\documentclass[11pt]{amsart}
\usepackage{geometry}                
\usepackage{graphicx}
\usepackage{amssymb}
\usepackage{amscd}
\usepackage{epstopdf}

\newtheorem{df}{Definition}[section]

\newtheorem{thm}{Theorem}[section]
\newtheorem{prop}{Proposition}[section]
\newtheorem{lm}{Lemma}[section]

\newtheorem{remark}{Remark}[section]
\newtheorem{fact}{Fact}[section]
\newtheorem{cor}{Corollary}[section]

\title{An MDS code associated to an elliptic curve}

\author{Ken-ichi Sugiyama
}

\begin{document}
\maketitle

\begin{center}
Department of Mathematics and Informatics,\\
 Faculty of Science, Chiba University.\\
 1-33 Yayoi-cho Inage-ku, Chiba\\
  263-8522, Japan\\
  e-mail address : sugiyama@math.s.chiba-u.ac.jp
\end{center}
\begin{abstract} We will construct an MDS($=$ the most distance separable)  code $C$ which admits a decomposition $C=\oplus_iC_i$ such that every factor is still MDS. An effective way of decoding will be also discussed.
\end{abstract}
\section{Introduction}
Let $q$ be a power of a prime $p$.  {\it A code} ${\mathcal C}$ is an imbedding
\begin{equation} e : C  \to {\mathbb F}_q^N,\quad N\geq 1,\end{equation}
where $C$ is a vector space over ${\mathbb F}_q$ and the minimal distance $d({\mathcal C})$ is defined to be one of $e(C)$, that is 
\[d({\mathcal C}):={\rm Min}_{x\neq 0 \in e(C)}w(x),\]
where $w(x)$ is {\it the weight} of the cordwood $x=(x_1,\cdots,x_N)$, i.e. the number of non-zero $x_i$. It is known $d({\mathcal C})\leq N-{\rm dim}\,C+1$ by the Singleton bound and if the equality hold the code is called {\it MDS}\footnote[1]{the Most Distance Separable} (\cite{M-S} {\bf Chapter 11}). If $d({\mathcal C}) \geq N-{\rm dim}\,C$, we mention ${\mathcal C}$ as {\it NMDS}\footnote[2]{Nearly the Most Distance Separable} .\\

A MDS or NMDS code has been naturally appeared in various references. Let $E$ be an elliptic curve defined over ${\mathbb F}_q$ and $\Sigma=\{s_1,\cdots,s_N\}$ a subset of $E({\mathbb F}_q)$, the set of ${\mathbb F}_q$-rational points. We identify the space ${\mathbb F}_q^{\Sigma}$ of ${\mathbb F}_q$-valued functions on $\Sigma$ with ${\mathbb F}_q^N$ by
\[{\mathbb F}_q^{\Sigma} \simeq {\mathbb F}_q^N, \quad f \to (f(s_1),\cdots,f(s_N)).\]
Let $D$ be a non-zero effective divisor on $E$ defined over ${\mathbb F}_q$ whose support  ${\rm Spt}(D)$ is disjoint from $\Sigma$. Then we have {\it the evaluation map},
\begin{equation}e : {\mathcal L}(D) \to {\mathbb F}_q^{\Sigma},\quad e(f)(s)=f(s)\quad (s\in \Sigma).\end{equation}
Here ${\mathcal L}(D):=\{f \in {\mathbb F}_q(E)\, |\, {\rm div}(f)+D \geq 0\}$, where ${\mathbb F}_q(E)$ is the space of rational functions on $E$ over ${\mathbb F}_q$. According to a situation,  we will often denote it by $H^0(E,{\mathcal O}(D))$, which is a notation of algebraic geometry. The map $e$ is injective if $N > {\rm deg}(D)$ and we assume it in the following. Then (2) is an NMDS code with ${\rm dim}{\mathcal L}(D)={\rm deg}(D)$ and is denoted by ${\mathcal C}_L(D,\Sigma)$. It is MDS iff for every subset $X$ of $\Sigma$ with $|X|={\rm deg}(D)$ the divisor $D-(X)$ is not principal, where $(X):=\sum_{x\in X}x \in {\rm Div}(E)$ and $|\cdot|$ is the cardinality(\cite{Shokrollahi}).  Fix $P\in \Sigma$ and set $\Sigma^{\ast}:=\Sigma\setminus P$. We define ${\mathcal L}_0(D)$ to be 
\[{\mathcal L}_0(D)=\{f\in {\mathcal L}(D)\,|\, f(P)=0\}.\]
Since the space of constant functions ${\mathbb F}_q$ is contained in ${\mathcal L}(D)$, ${\rm dim}{\mathcal L}_0(D)={\rm dim}{\mathcal L}(D)-1$ and one see that the code ${\mathcal C}^0_L(D,\Sigma^{\ast}) $ defined by 
\[e^{\ast} : {\mathcal L}_0(D) \to {\mathbb F}_q^{\Sigma^{\ast}},\quad e(f)(s)=f(s) \quad (s\in \Sigma^{\ast}).\]
is NMDS. Moreover it is MDS if so is ${\mathcal C}_L(D,\Sigma)$ (cf. {\bf Lemma 2.1}). \\

Suppose that the code (1) is NMDS (resp. MDS). A decomposition $C=\oplus_i C_i$ is mentioned as {\it proper} if the restriction
\[e : C_i \to {\mathbb F}^N,\]
is also NMDS (resp. MDS) for each $i$.

Set $k={\mathbb F}_q$ and $k_m={\mathbb F}_{q^m}$ for a positive integer $m$. Fix a positive square free integer $N$ and let ${\mathcal D}(N)$ describe the set of divisors of $N$. Let $E$ be an elliptic curve defined over $k$. Take a prime $l\neq p$ and we consider a representation
\[\rho_{l} : {\rm End}_{k}(E) \to {\rm End}\,T_l(E),\]
where $T_{l}(E)$ is the $l$-adic Tate module. Let $F$ be the $q$-th power endomorphism of $E$ and $\alpha$ an eigenvalue of $\rho_{l}(F)$, which is an algebraic integer with modulus $\sqrt{q}$. For a positive integer $n$ we put
\[(x)_{n}=\sum_{i=0}^{n-1}x^i=\frac{x^n-1}{x-1},\]
and denote the set of primes dividing $n$ by $P(n)$. 
\begin{thm} Let $E$ be an elliptic curve over $k$ and $N$ a positive square free integer. Then there are  effective divisors $D_N$ and $\{D_{r}\}_{r\in P(N)}$ defined over $k_N$ which satisfies the following properties.
\begin{enumerate}
\item 
\[D_r  \leq D_N, \quad \forall r \in P(N).\]
\item There is a decomposition
\[{\mathcal L}_0(D_N)=\oplus_{r\in P(N)}{\mathcal L}_0(D_r),\]
and the projector to the $r$-th factor $\phi_{N/r}^0$ is explicitly described.
\item 
\[{\rm dim}{\mathcal L}_0(D_r)=|(\alpha)_r|^2-1, \quad \forall r\in P(N),\] 
and in particular 
\[{\rm dim}{\mathcal L}_0(D_N)=\sum_{r\in P(N)}\{|(\alpha)_r|^2-1\}.\]
 \end{enumerate}
\end{thm}
In order to construct a code we take a finite subset $\Sigma$ of $E$ disjoint from the support of $D_N$ with $|\Sigma| > {\rm deg}(D_N)$. In fact we can take such a subset in $E(k_N)\setminus {\rm Spt}(D_N)$ (cf. {\bf Lemma 4.1}). Then ${\mathcal C}^{0}_L(D_N, \Sigma^\ast)$ is automatically NMDS  defined over $k_N$ and the decomposition of {\bf Theorem1.1} (2) is proper. In order to obtain an MDS code which admits a proper decomposition the construction of $\Sigma$ is rather involved.
\begin{thm} Let $D_N$ and $\{D_{r}\}_{r\in P(N)}$ be effective divisors in {\bf Theorem 1.1} and take an arbitrary integer $m$ greater than ${\rm deg}(D_N)$. Then there is a subset $\Sigma$ of $E(k_N^\prime)$ ($k_N^\prime$ is a finite extension of $k_N$)  which satisfies the following properties.
\begin{enumerate}
\item $\Sigma$ is disjoint from the support of $D_N$ and $|\Sigma|=m$. 
\item All the codes ${\mathcal C}_L^0(D_N,\Sigma^\ast)$ and $\{{\mathcal C}_L^0(D_r,\Sigma^\ast)\}_{r\in P(N)}$ are MDS. In particular the decomposition
\[{\mathcal L}_0(D_N)=\oplus_{r\in P(N)}{\mathcal L}_0(D_r),\]
is proper. 
\end{enumerate}
\end{thm}
We will show a concrete example of {\bf Theorem 1.2} in the final section(cf. {\bf Theorem 6.1}). In general it is not hard to construct a code that admits a proper decomposition from a divisor on an elliptic curve (cf. {\bf Proposition 2.1} and {\bf Proposition 2.2}). But in order to describe $\phi_{N/r}^0$ explicitly we will construct $D_N$ and $\{D_r\}_{r\in P(N)}$ from the kernel of an isogeny of the elliptic curve (cf. {\bf Theorem 4.3}). \\

Here is a significance of the theorem. Take {\it a word} $w\in {\mathcal L}_0(D_N)$ and let $c:=e^\ast(w) \in k_N^{\Sigma^\ast}$ be the corresponding code word. We transmit $c$ and let $c^\prime$ be the received vector. Then because of an interference  $c^\prime=c+\epsilon,$ where $\epsilon$ is an error. If the weight of $\epsilon$ is less than the half of the minimal distance of ${\mathcal C}_L({\mathcal L}_0(D_N),\Sigma^\ast)$ we can correct errors by the Pellikaan's algorithm (see \$5). Our aim is to find an another way which may correct an error of a larger weight. First we decompose $w=\sum_{r \in P(N)}\phi_{N/r}^0(w)$ and we will use the family $\{c_r:=e^\ast(\phi_{N/r}^0(w))\}_r$ as a code word. Transmit them and let $\{c^\prime_r\}_r$ be the received vectors. As before by the Pellikaan's algorithm we can properly decode $c^\prime_r$ if the weight of the error vector is less than the half of $d({\mathcal C}_L({\mathcal L}_0(D_r),\Sigma^\ast))$. Sum them up and then we will recover the original word $w$ if the weight of the error of $c_r^\prime$ is less than the half of ${\rm Min}_r d({\mathcal C}_L({\mathcal L}_0(D_r),\Sigma^\ast))$ for every $r$. Since ${\rm Min}_r d({\mathcal C}_L({\mathcal L}_0(D_r),\Sigma^\ast))> d({\mathcal C}_L({\mathcal L}_0(D_N),\Sigma^\ast))$ we expect that the latter method may correct more errors than the previous (i.e. usual) one. \\

\section{A proper decomposition}
Let $E$ be an elliptic curve defined over a finite field $k$ of characteristic $p$. In the following we fix a subset $\Sigma$ of $E(k)$ and a point $P\in \Sigma$.
\begin{lm} Let $D$ be a nonzero effective divisor whose support is disjoint from $\Sigma$. If $|\Sigma| > {\rm deg}(D)$, ${\mathcal C}_L^0(D,\Sigma^{\ast})$ is NMDS. If moreover ${\mathcal C}_L(D,\Sigma)$ is MDS, so is ${\mathcal C}^0_L(D,\Sigma^\ast)$.
\end{lm} 
{\bf Proof.}  We first claim that $e^{\ast} : {\mathcal L}_0(D) \to k^{\Sigma^{\ast}} $ is injective. We define
\[e_P : {\mathcal L}(D) \to k,\quad e_P(s)=s(P),\]
and 
\[\nu_P : k^{\Sigma} \to k,\quad \nu_P(f)=f(P),\]
and let $\epsilon : k^{\Sigma^{\ast}} \to k^{\Sigma}$ be the extension by $0$ on $P=\Sigma\setminus \Sigma^{\ast}$. Then 
\[\begin{CD}
 0 @>>> {\mathcal L}_0(D) @>>>  {\mathcal L}(D) @>\text{$e_P$}>> k @>>>0\\
@VVV @V\text{$e^{\ast}$}VV  @V\text{$e$}VV @V\text{${\rm id}$}VV @VVV \\
0@>>>  k^{\Sigma^{\ast}} @>\text{$\epsilon$}>>  k^{\Sigma} @>\text{$\nu_P$}>> k @>>> 0,
 \end{CD}\]
shows that if $e$ is injective so is $e^\ast$. Hence the assumption implies the claim. Next we investigate the minimal distance. For a function $f$ on $\Sigma$ we set $\nu(f):=|\Sigma|-w(f)$, that is the number of zeros. Let $f\neq 0 \in  {\mathcal L}_0(D)$ and we have to show that
 \[w(e^{\ast}(f)) \geq |\Sigma^{\ast}|-{\rm dim} {\mathcal L}_0(D)=|\Sigma|-{\rm deg} (D).\]
 Suppose that $w(e^{\ast}(f))<|\Sigma|-{\rm deg} (D)$. Since $|\Sigma|-w(e^{\ast}(f))=\nu(e(f))$, this is equivalent to $\nu(e(f)) >  {\rm deg} (D)$ and $f=0$. Hence ${\mathcal C}^0_L(D,\Sigma^\ast)$ is NMDS. Finally suppose that ${\mathcal C}_L(D,\Sigma)$ be an MDS code and let $g\in {\mathcal L}(D)$. Then $\nu(e(g)) \geq {\rm deg}(D)$ implies $g=0$. If $f\in {\mathcal L}_0(D)$ satisfies $w(e^\ast(f)) \leq |\Sigma^{\ast}|-{\rm dim} {\mathcal L}_0(D)$, $\nu(e(f)) \geq {\rm deg}(D)$ by the above equations and $f=0$. Therefore ${\mathcal C}^0_L(D,\Sigma)$ is MDS.
\begin{flushright}
$\Box$
\end{flushright}

\begin{lm}
Let $X$ and $Y$ be subsets of $E(k)$ that are disjoint from $\Sigma$. Suppose that they meet at a single point. Then
\[{\mathcal L}_0((X\cup Y))={\mathcal L}_0((X))\oplus {\mathcal L}_0((Y)).\]
\end{lm}
{\bf Proof.} 
Let $Q$ be the intersection of $X$ and $Y$. Then
\[H^{0}(E,{\mathcal O}(X))\cap H^{0}(E,{\mathcal O}(Y))=H^{0}(E,{\mathcal O}(Q))= k,\]
which yields an exact sequence
\[0 \to k \to H^{0}(E,{\mathcal O}(X))\oplus  H^{0}(E,{\mathcal O}(Y)) \stackrel{s}\to H^{0}(E,{\mathcal O}(X\cup Y)),\]
where $s(f,g)=f+g$. Since
\[{\rm dim}H^{0}(E,{\mathcal O}(X\cup Y))=|X|+|Y|-1={\rm dim}H^{0}(E,{\mathcal O}(X))+{\rm dim}H^{0}(E,{\mathcal O}(Y))-1,\]
$s$ is surjective. For a rational function $f$ regular at $P$ we define $e_P(f):=f(P)$ as before and set
\[\delta(x)=(x,x),\quad \sigma(x,y)=x+y, \quad x,y\in k.\]
Then take the kernels of the vertical arrows of the following diagram and the claim is obtained:
\[\begin{CD}
 0@>>>  k @>>> H^{0}(E,{\mathcal O}(X))\oplus  H^{0}(E,{\mathcal O}(Y)) @>\text{$s$}>> H^{0}(E,{\mathcal O}(X\cup Y)) @>>> 0\\
 @VVV  @V\text{${\rm id}$}VV @V\text{$e_P\oplus e_P$}VV @V\text{$e_P$}VV @VVV\\
  0 @>>> k@ >\text{$\delta$}>> k\oplus k  @>\text{$\sigma$}>> k @>>> 0.
 \end{CD}\]
\begin{flushright}
$\Box$
\end{flushright}
\begin{prop} Let $\{X_i\}_i$ be a finite family of subsets of $E(k)$ disjoint from $\Sigma$ which meet at a single point. Then
\[{\mathcal L}_0((\cup_i X_i))=\oplus_i {\mathcal L}_0((X_i)).\] 
\end{prop}
In general let $\{Y_i\}_{i\in I}$ be a finite family of subsets of $E(k)$ and $m$ a positive integer greater than $|Y|$ where $Y:=\cup_{i \in I}Y_i$.
\begin{prop} There is a subset $\Sigma$ of $E(k^\prime)$ where $k^\prime$ is a finite extension of $k$ which satisfies the following properties.
\begin{enumerate}
\item 
\[|\Sigma|=m, \quad \Sigma \cap Y =\phi.\]
\item ${\mathcal C}_L((Y_i),\Sigma)$ is MDS for all $i\in I$.
\end{enumerate}
\end{prop}
{\bf Proof.}  For a positive integer $t$ less than $m$, let ${\mathcal J}_t$ be the collection of subsets of $\{1,\cdots,m\}$ with cardinality $t$ and we associate the projection $\pi_J$ with $J\in {\mathcal I}_t$ by
\[\pi_J : E^m \to E^t,\quad \pi_I(x_1,\cdots,x_m)=(x_j)_{j \in J}.\]
Put $d_i=|Y_i|$ and we define an epimorphism $\sigma_i : E^{d_i} \to E$ to be
\[\sigma_i(x_1,\cdots,x_{d_i})=\sum_{j=1}^{d_i} x_j-\sum_{y\in Y_i}y.\]
Let $\sigma_J$ be the composition $\sigma_J:=\sigma_i\cdot \pi_J$ for $J\in {\mathcal J}_{d_i}$. We define the divisors $Z$, $W$ and $\Delta$ of $E^m$ as follows:
\begin{enumerate}
\item 
\[W:=\cup_{i=1}^m\pi_i^{-1}(Y),\]
where $\pi_i$ is the projection to the $i$-th factor. 
\item
\[Z:=\cup_{i\in I}\cup_{J\in {\mathcal J}_{d_i}}\sigma_J^{-1}(0).\]
\item 
\[\Delta:=\{(x_1,\cdots,x_m)\in E^m\,|\, x_i=x_j \quad (\exists i\neq j)\}.\]
\end{enumerate}
Take $x=(x_1,\cdots,x_{m})\in E(k^\prime)^{m}\setminus (Z\cup W \cup \Delta)$ and we define a subset $\Sigma$ of $E$ as $\Sigma:=\{x_1,\cdots,x_m\}$. Here note that such $x$ exists if $k^\prime$ is sufficiently large. The condition (1) implies that $x_i \notin Y$ for all $i$ and $\Sigma \cap Y=\phi$. We find that $|\Sigma|=m$ by (3). The condition (2) shows that $(Y_i)-(\sum_{j\in J} x_j)$ is not principal for $\forall J\in {\mathcal J}_{d_i}$ ($\forall i\in I$). Hence ${\mathcal C}_L((Y_i),\Sigma)$ is MDS for all $i\in I$ as we have explained in the introduction.
\begin{flushright}
$\Box$
\end{flushright}
The following theorem is clear from {\bf Lemma 2.1}, {\bf Proposition 2.1} and {\bf Proposition 2.2}.
\begin{thm} Let $\{X_i\}_i$ be a finite family of subsets of $E(k)$ which meet at a single point. Set $X:=\cup_iX_i$ and let $m$ be an arbitrary integer greater than $|X|$. Then there is a subset $\Sigma$ of $E(k^\prime)\setminus X$ where $k^\prime$ is a finite extension of $k$ such that
\begin{enumerate}
\item $|\Sigma|=m$.
\item ${\mathcal C}^0_L((X),\Sigma^\ast)$ and ${\mathcal C}^0_L((X_i),\Sigma^\ast)\, (\forall i)$ are MDS. 
\item ${\mathcal L}_0((\cup_i X_i))$ has a proper decomposition,
\[{\mathcal L}_0((X))=\oplus_i {\mathcal L}_0((X_i)).\] 
\end{enumerate}
\end{thm}
In order to describe the projector explicitly we impose a certain structure of $\{X_i\}_i$, which will be discussed in the following sections.
\section{The kernel of an isogeny}
Let $q$ be a power of a prime $p$ and set $k={\mathbb F}_q$ and $k_m={\mathbb F}_{q^m}$. Let $E$ be an elliptic curve defined over $k$. Take a prime $l\neq p$ and we consider a representstion
\[\rho_{l} : {\rm End}_{k}(E) \to {\rm End}\,T_l(E),\]
where $T_{l}(E)$ is the $l$-adic Tate module. We recall facts which will be used later.
\begin{fact} \cite{Silverman}
\begin{enumerate}
\item 
\[{\rm deg}\,f={\rm det}\rho_{l}(f), \quad f\in {\rm End}_{k}(E).\]
\item Let $F$ be the $q$-th power endomorphhism of $E$ and $\{\alpha,\beta\}$ eigenvalues of $\rho_{l}(F)$. Then they are algebraic integers with modulus $\sqrt{q}$ and are mutually complex conjugate.
\item 
\[|E(k_m)|=(1-\alpha^m)(1-\beta^m)=|1-\alpha^m|^2.\]
\end{enumerate}
\end{fact}
For positive integers $m$ and $n$ we define an endomorphism $\tau_{mn/m}$ of $E$ by
\[\tau_{mn/m}=\sum_{i=0}^{n-1}F^{im}=\frac{1-F^{mn}}{1-F^m}.\]
Then
\begin{equation}
\tau_{mn/1}=\frac{1-F^{mn}}{1-F}=\frac{1-F^{mn}}{1-F^m}\frac{1-F^m}{1-F}=\tau_{mn/m}\cdot \tau_{m/1}.
\end{equation}
Let $G_{m}$ be the kernel of $\tau_{m/1}$. Since the differential of $\tau_{m/1}$ is equal to $1$, $\tau_{m/1}$ is separable and $G_{m}$ is a reduced subgroup of $E$. In particular ${\rm deg}(\tau_{m/1})=|G_m|$.
\begin{lm}
\begin{enumerate}
\item If $m|n$, $G_m \subset G_n$.
\item The order of $G_{m}$ is $|(\alpha)_{m}|^2$.
\item $G_{m}$ is contained in $E(k_m)$ and
\[0 \to G_{m} \to E(k_m) \stackrel{\tau_{m/1}} \to E(k) \to 0.\]
\end{enumerate}
\end{lm}
{\bf Proof.}(1) follows from the equation (3).\\
(2) Since $|G_{m}|={\rm deg}\tau_{m/1}$ {\bf Fact 3.1} (1) and (2) show
\[|G_{m}|={\rm det}\rho_{l}(\tau_{m/1})=(\sum_{i=0}^{m-1}\alpha^{i})(\sum_{i=0}^{m-1}{\beta}^{i})=|(\alpha)_{m}|^2.\]
(3) Take $x\in G_m$ and
\[(F^m-1)(x)=(F-1)\tau_{m/1}(x)=0.\]
Thus $F^{m}(x)=x$, which shows $G_{m} \subset E(k_m)$. On the other for $y\in E(k_m)$,
\[(F-1)\tau_{m/1}(y)=(F^m-1)(y)=0,\]
and $\tau_{m/1}(y) \in E(k)$. Thus we have an exact sequence
\[0 \to G_{m} \to E(k_m) \stackrel{\tau_{m/1}} \to E(k).\]
Since by {\bf Fact 3.1}(3), $|E(k_m)|=|1-\alpha^m|^2=|(\alpha)_m|^2 |1-\alpha|^2=|G_m|\cdot |E(k)|$ and $\tau_{m/1}$ is surjective.
\begin{flushright}
$\Box$
\end{flushright}
\begin{remark} From the proof, we see that the degree of $\tau_{m/1}$ is $|(\alpha)_m|^2$, which is prime to $p$ by {\bf Fact 3.1}(2).
\end{remark}
\begin{lm} Suppose that $gcd(m,n)=1$. Then
\[G_m\cap G_n=0.\]
\end{lm}
{\bf Proof.} Since $m$ and $n$ are coprime $E(k_m)\cap E(k_n) =E(k)$ and $G_m\cap G_n \subset E(k)$ by {\bf Lemma 3.1}(3).
Take $x\in G_m\cap G_n$. Because  $F(x)=x$, 
\[0=\tau_{m/1}(x)=\sum_{i=0}^{m-1}F^{i}(x)=mx,\]
and similarly $nx=0$.
Since $gcd(m,n)=1$, $x=0$.
\begin{flushright}
$\Box$
\end{flushright}
\begin{prop} Let $m$ and $n$ be positive coprime integers. Then
\[{\rm Ker}(\tau_{m/1}\tau_{n/1})=G_{m}+G_{n}.\]
\end{prop}
\begin{remark}
In fact by {\bf Lemma 3.2} RHS is a direct sum.
\end{remark}
{\bf Proof.} Since $\tau_{m/1}$ and $\tau_{n/1}$ commute, 
\[G_{m}+G_{n} \subset {\rm Ker}(\tau_{m/1}\tau_{n/1}).\]
We compare the orders of both sides. By {\bf Lemma 3.2}, $|G_{m}+G_{n}|=|G_{m}|\cdot |G_{n}|$ and  
\[|G_{m}+G_{n}|={\rm deg}(\tau_{m/1}){\rm deg}(\tau_{n/1})={\rm deg}(\tau_{m/1}\tau_{n/1})=|{\rm Ker}(\tau_{m/1}\tau_{n/1})|,\]
which implies the claim.
\begin{flushright}
$\Box$
\end{flushright}
\section{A construction of a code}
We fix a positive square free integer $N$ and consider the base extension of $E$ to ${\rm Spec}\, k_N$, which will be denoted by the same letter. We describe the set of positive divisors of $N$ by ${\mathcal D}(N)$. For $m\in {\mathcal D}(N)$ we set 
\[X_m:=\cup_{r\in P(m)}G_r, \quad D_m:=(X_m). \]
{\bf Lemma 3.1} implies that $X_m$ is contained in $G_m$ and that $D_m$ is defined over $k_m$. The following proposition is clear from {\bf Lemma 3.1} and {\bf Lemma 3.2}.
\begin{prop} 
\[{\rm deg}\,(D_m)-1=\sum_{r\in P(m)}\{{\rm deg}(G_r)-1\}=\sum_{r\in P(m)}\{|(\alpha)_r|^2-1\}.\]
\end{prop}

\begin{df} For $m\in {\mathcal D}(N)$ we define an endomorphism $\pi_m$ of $E$ to be
\[\pi_{m}=\prod_{r \in P(m)} \tau_{r/1}.\]
\end{df}
The degree of $\pi_{m}$ is $\prod_{r \in P(m)} {\deg}(\tau_{r/1})$, which is prime to $p$ (cf. {\bf Remark 3.1}). If $m$ and $n$ are coprime, 
$\pi_{m}\cdot \pi_{n}=\pi_{mn}=\pi_{n}\cdot \pi_{m}$.
\begin{prop} Let $m,n \in {\mathcal D}(N)$ be coprime.
\begin{enumerate}
\item
\[{\rm Ker}(\pi_m)=\oplus_{r\in P(m)}G_{r}.\]
\item
\[0 \to {\rm Ker}(\pi_m) \to {\rm Ker}(\pi_{mn}) \stackrel{\pi_m}\to {\rm Ker}(\pi_{n}) \to 0.\]
\item
\[0 \to {\rm Ker}(\pi_m)\to E(k_{mn}) \stackrel{\pi_m}\to E(k_n) \to 0.\]
\item
\[X_m \subset {\rm Ker}(\pi_m).\]
\item 
\[\pi_{m}(X_{mn})=X_n.\]
\end{enumerate}
\end{prop}
{\bf Proof.}
(1) and (2) follow from {\bf Proposition 3.1}. Using {\bf Lemma 3.1}(3) successively one obtain (3). (4) and (5) are the consequences of (1) and (2).
\begin{flushright}
$\Box$
\end{flushright}
From {\bf Proposition 2.1} and {\bf Lemma 3.2}, we obtain the following theorem.
\begin{thm}
\[{\mathcal L}_0(D_N)=\oplus_{r\in P(N)}{\mathcal L}_0(D_r).\]
\end{thm}
\begin{lm}If $|E(k)|\geq 2$,
\[|E(k_N)\setminus X_N| > {\rm deg}(D_N)(=|X_N|).\]
\end{lm}

\begin{remark} {\bf Fact 3.1}(2) and (3) imply that the assumption is satisfied if $|k| \geq 5$.
\end{remark}
{\bf Proof.}
By  {\bf Proposition 4.2} we see that 
\[X_N \subsetneqq G_N:=\tau_{N/1}^{-1}(0),\quad E(k_N)\setminus G_N \subsetneqq E(k_N)\setminus X_N.\]
On the other hand since by {\bf Lemma 3.1}(3), $E(k_N)\setminus G_N=(\tau_{N/1})^{-1}(E(k)\setminus 0)$ and the assumption yields
\[ |E(k_N)\setminus G_N| \geq |G_N|.\,\]
and the claim is clear.
\begin{flushright}
$\Box$
\end{flushright}

Therefore there is a subset $\Sigma$ of $E(k_N)\setminus {\rm Spt (D_N)}$ such that $|\Sigma| > {\rm deg}(D_N)$ (e.g. $\Sigma:=E(k_N)\setminus X_N$). Hence by {\bf Lemma 2.1}, ${\mathcal C}_L^0(D_N,\Sigma^\ast)$ is NMDS and the decomposition of {\bf Theorem 4.1} is proper.

\begin{thm} Let $m$ be an arbitrary integer greater than ${\rm deg}(D_N)$. Then there is a subset $\Sigma$ of $E(k_N^\prime)\setminus {\rm Spt}(D_N)$ where $k_N^\prime$ is a finite extension of $k_N$ such that 
\begin{enumerate}
\item $|\Sigma|=m$.
\item ${\mathcal C}^0_L(D_N,\Sigma^{\ast})$ is an MDS code and the decomposition of {\bf Theorem 4.1} is proper.
\end{enumerate}
\end{thm}
{\bf Proof.}
The statement is clear from {\bf Theorem 2.1} and {\bf Lemma 3.2}.
\begin{flushright}
$\Box$
\end{flushright}
Let $m, n \in {\mathcal D}(N)$ be coprime. Since $D_m \leq  D_{mn}$, $H^{0}(E,{\mathcal O}(D_m))$ is a subspace of $H^{0}(E,{\mathcal O}(D_{mn}))$ and let $\iota$ be the inclusion. We denote the composition of
\[H^{0}(E,{\mathcal O}(D_m)) \stackrel{\iota}\to H^{0}(E,{\mathcal O}(D_{mn})) \stackrel{(\pi_n)_{\ast}}\to H^{0}(E,{\mathcal O}(D_m)),\]
by $p_{mn/m}$. Here the latter map is obtained by {\bf Proposition 4.2}(5).
\begin{prop}
\begin{enumerate}
\item If $f\in H^{0}(E,{\mathcal O}(D_n))$, $(\pi_n)_{\ast}(f)$ is a constant.
\item $p_{mn/m}$ is an isomorphism so that
\[p_{mn/m}(1)={\rm deg}(\pi_n)\neq 0.\]
\end{enumerate}
\end{prop}
{\bf Proof.} (1) By {\bf Proposition 4.2}(4) $\pi_n(X_n)=0$ and
\[(\pi_n)_{\ast}(f) \in H^{0}(E,{\mathcal O}(0))=k_N.\]
(2) It is sufficient to show that $p_{mn/m}$ is injective. For a rational function $f$ on $E$ let ${\mathcal P}(f)$ describe the set of poles.  By definition
\[p_{mn/m}(f)(y)=\sum_{z\in \pi_n^{-1}(y)}f(z)=\sum_{\gamma \in {\rm Ker}(\pi_n)}f(x+\gamma),\]
where $\pi_n(x)=y$. In particular $p_{mn/m}(1)={\rm deg}(\pi_n)$. Suppose $f\in H^{0}(E,{\mathcal O}(D_m))$ satisfies that $p_{mn/m}(f)=0$. Then
\begin{equation}f(x)=-\sum_{\gamma\neq 0 \in {\rm Ker}(\pi_n)}f(x+\gamma).\end{equation}
Since, by {\bf Proposition 4.2}(1), the support ${\rm Spt}{\mathcal P}(f)$ is contained in ${\rm Ker}(\pi_m)$, ${\rm Spt}{\mathcal P}(f_{\gamma})\subset {\rm Ker}(\pi_m)+\gamma$. Thus (5) shows
\[{\rm Spt}{\mathcal P}(f) \subset \cup_{\gamma\neq 0\in {\rm Ker}(\pi_n)}\{{\rm Ker}(\pi_m)\cap \{{\rm Ker}(\pi_m)+\gamma\}\}.\]
But ${\rm Ker}(\pi_m)\cap \{{\rm Ker}(\pi_m)+\gamma\}=\phi$ for $\gamma\neq 0\in {\rm Ker}(\pi_n)$ because ${\rm Ker}(\pi_m)\cap {\rm Ker}(\pi_n)=0$ by {\bf Lemma 3.2} (Here note that ${\rm Ker}(\pi_m) \subset G_{m}$ by {\bf Lemma 3.1}(1) and {\bf Proposition 4.2}(1)). Thus $f$ should be a constant and 
\[p_{mn/m}(f)={\rm deg}(\pi_{n})f.\]
Since ${\rm deg}(\tau_{m/1})$ is prime to $p$ as we have mentioned in {\bf Remark 3.1} so is ${\rm deg}(\pi_{m})$ and we obtain the claim. 
\begin{flushright}
$\Box$
\end{flushright}
For $r\in P(N)$ we define 
\[\phi_{N/r} : H^{0}(E,{\mathcal O}(D_{N})) \to H^{0}(E,{\mathcal O}(D_r)), \quad \phi_{N/r}:=(p_{N/r})^{-1}\cdot (\pi_{N/r})_{\ast}.\]
Then by definition,
\begin{equation}\phi_{N/r}(h)=h, \quad h\in  H^{0}(E,{\mathcal O}(D_r)),\end{equation}
and if $g\in H^{0}(E,{\mathcal O}(D_{r^\prime}))$ $(r^\prime \neq r \in P(N))$, $\phi_{N/r}(g)$ is a constant by {\bf Proposition 4.3}.

Set
\[\phi_{N/r}^0(f):=\phi_{N/r}(f)-\phi_{N/r}(f)(P), \quad  f\in H^{0}(E,{\mathcal O}(D_{N})).\]
\begin{thm} $\phi_{N/r}^0$ is a linear map from $H^{0}(E,{\mathcal O}(D_{N}))$ to ${\mathcal L}_0(D_r)$ satisfying the following properties.
\begin{enumerate}
\item 
\[\phi_{N/r}^0(h)=h, \quad h\in  {\mathcal L}_0(D_r).\]
\item If $r^\prime\neq r$,
\[\phi_{N/r}^0(h)=0, \quad h\in  {\mathcal L}(D_{r^\prime}).\]
\item For $f\in {\mathcal L}_0(D_N)$,
\[ f=\sum_{r\in P(N)}\phi_{N/r}^0(f).\]
\end{enumerate}
\end{thm}
{\bf Proof.} (1) and (2) are clear from the definition. By {\bf Theorem 4.1}
we may describe $f=\sum_{s\in P(N)}f_s$, $f_s\in {\mathcal L}_0(D_s)$. Use (1) and (2), and
\[\phi_{N/r}^0(f)=\sum_{s\in P(N)}\phi_{N/r}^0(f_s)=f_r,\]
which shows (3).
\begin{flushright}
$\Box$
\end{flushright}
Now {\bf Theorem 1.1} and {\bf Theorem 1.2} follow from {\bf Proposition 4.1}, {\bf Theorem 4.1}, {\bf Theorem 4.2} and {\bf Theorem 4.3}.
\section{Error correcting pairs}
Let $k$ be a finite field of characteristic $p$. For a finite set $\Sigma=\{s_1,\cdots,s_n\}$, we identify $k^\Sigma$ with $k^n$ as the introduction. The bilinear form $(\cdot,\cdot)$ on $k^n$ is defined by $({\mathbf x},{\mathbf y})=\sum_{i}x_iy_i$. If $V$ is a linear subspace of $k^n$ let $V^{\perp}$ be the orthogonal complement; $V^{\perp}=\{{\mathbf x}\in k^\Sigma\,|\, ({\mathbf x},{\mathbf v})=0, \forall {\mathbf v}\in V\}$. For ${\mathbf x}, \,{\mathbf y}\in k^n$ {\it the star multiplication} ${\mathbf x}\ast{\mathbf y}\in k^\Sigma$ is defined by the coordinate-wise multiplication, that is $({\mathbf x}\ast{\mathbf y})_i=x_iy_i$. For two subset $A$ and $B$ of $k^n$ we denote the set $\{{\mathbf a}{\ast}{\mathbf b}\,|\,{\mathbf a}\in A,\,{\mathbf b}\in B\}$ by $A{\ast}B$. Now let $e : C \to {\mathbb F}^n$ be a code.
\begin{df} {\rm A $t$-error correcting pair} $(A,B)$ for $C$ is defined to be a pair of linear subspace $A$ and $B$ of $k^n$ satisfying the following conditions:
\begin{enumerate}
\item \[A{\ast}B \subset e(C)^{\perp}.\]
\item \[{\rm dim}(A) >t.\]
\item
\[d(A)+d(C)>n.\]
\item \[d(B^{\perp}) >t.\]
\end{enumerate}
\end{df}
\begin{fact}(\cite{Pellikaan} {\bf Theorem 2.14}) If $(A,B)$ is a $t$-error correcting pair for $C$, there is an effective algorithm which corrects $t$-errors with complexity $O(n^3)$. 
\end{fact}
For the actual algorithm see \cite{Pellikaan} {\bf Algorithm  2.13}. We will apply the {\bf Fact 4.1} to our code. Let $N$ be a positive square free integer and $m\in{\mathcal D}(N)$. Let $\Sigma$ be a finite subset of $E$ disjoint from ${\rm Spt}(D_N)$ with $|\Sigma|>{\rm deg}(D_N)$. We choose an arbitrary point $P$ of $\Sigma$ and set $\Sigma^{\ast}:=\Sigma\setminus P$ as before. Embed ${\mathcal L}_0(D_m)$ into ${\mathcal L}(D_m)$ and we will find $d^{\ast}$-error correcting pair for ${\mathcal C}_L(D_m, \Sigma^\ast)$. Here
\[d^{\ast}:=\lfloor \frac{1}{2}(|\Sigma^{\ast}|-{\rm deg}(D_m))\rfloor -1,\]
where  $\lfloor x \rfloor$ is the maximal integer less than or equal to $x$. Since $d({\mathcal C}_L(D_m, \Sigma^\ast)) \geq |\Sigma^\ast|-{\rm deg}(D_m)$, $d^{\ast}$ is the maximal weight of errors which may be corrected.\\

In general let $\Sigma$ be a subset of $E(k)$. Let $D$ be an effective divisor defined over $k$ whose support is disjoint from $\Sigma$.  We denote the image of 
\[e : H^{0}(E,{\mathcal O}(D)) \to k^{\Sigma},\quad e(f)(x)=f(x),\]
and
\[r : H^{0}(E,\Omega^{1}((\Sigma)-D)) \to k^{\Sigma}, \quad r(\omega)(x)={\rm Res}_{x}(\omega),\]
by $C_{L}(D, \Sigma)$ and $C_{\Omega}((\Sigma)-D, \Sigma)$, respectively. Here ${\rm Res}_{x}(\omega)$ is the residue of $\omega$ at $x$. Since the canonical sheaf $\Omega$ is trivial, if we fix an invariant differential of $E$, $C_{\Omega}((\Sigma)-D, \Sigma)$ is identified with $C_{L}((\Sigma)-D, \Sigma)$. Assume that ${\rm deg}(D)<|\Sigma|$. Then the above maps are injective and the images are mutually orthogonal complement since the sum of all residues of a rational 1-form is zero.
\begin{prop} Suppose that $d^\ast$ is positive. Then the pair of $C_{L}( (d^{\ast}+1)(P), \Sigma^{\ast})$ and $C_{\Omega}((\Sigma^{\ast})-D_m-(d^{\ast}+1)(P), \Sigma^{\ast})$ is a $d^{\ast}$-error correcting pair for $C_{L}(D_m, \Sigma^{\ast})$.  
\end{prop}
{\bf Proof.} Note that the assumption implies that the evaluation maps, ${\mathcal L}(D_m)\stackrel{e^\ast}\to k^{\Sigma^\ast}$, ${\mathcal L}((d^\ast+1)P)\stackrel{e^\ast}\to k^{\Sigma^\ast}$ and ${\mathcal L}(D_m+(d^\ast+1)P)\stackrel{e^\ast}\to k^{\Sigma^\ast}$ are injective. Set
\[A:=C_{L}( (d^{\ast}+1)(P), \Sigma^{\ast}),\quad B:=C_{\Omega}((\Sigma^{\ast})-D_m-(d^{\ast}+1)(P), \Sigma^{\ast}),\]
and
\[C:=C_{L}(D_m, \Sigma^{\ast}),\]
and we will check whether the conditions of {\bf Definition 5.1} are satisfied. The star multiplication $A\ast B$ is contained in $C_{\Omega}((\Sigma^{\ast})-D_m, \Sigma^{\ast})$, which is the orthogonal complement of $C$. Hence (1) has been checked. Since ${\rm dim}(A)=d^{\ast}+1$, (2) is clear. For (3) note that
\[d(A) \geq |\Sigma^{\ast}|-(d^{\ast}+1),\quad d(C) \geq  |\Sigma^\ast| - {\rm deg}(D_m),\]
and 
\[d(A)+d(C)-|\Sigma^{\ast}| \geq |\Sigma^\ast| - {\rm deg}(D_m)- (d^{\ast}+1)\geq d^\ast+1 >0 .\]
Finally we check (4). Observe that
\[B^{\perp}=C_L(D_m+(d^{\ast}+1)(P), \Sigma^{\ast}),\]
and
\[d(B^{\perp}) \geq |\Sigma^\ast|-{\rm deg}(D_m)-(d^{\ast}+1) \geq \frac{1}{2}(|\Sigma^{\ast}| - {\rm deg}(D_m)) \geq  d^{\ast}+1.\]
\begin{flushright}
$\Box$
\end{flushright}
\begin{cor}The pair of $C_{L}( (d^{\ast}+1)(P), \Sigma^{\ast})$ and $C_{\Omega}((\Sigma^{\ast})-D_m-(d^{\ast}+1)(P), \Sigma^{\ast})$ is a $d^{\ast}$-error correcting pair for $e^\ast({\mathcal L}_0(D_m))$.
\end{cor}
\begin{remark} The minimal distance of ${\mathcal C}_L^0(D_m,\Sigma^{\ast})$ is greater than or equal to $|\Sigma|-{\rm deg}(D_m)$ and 
\[d^{\ast}=\lfloor \frac{1}{2}(|\Sigma|-{\rm deg}(D_m))\rfloor -1,\]
if $|\Sigma|-{\rm deg}(D_m)$ is odd.
\end{remark}
Thus we have an effective decoding algorithm of ${\mathcal C}_L^0(D_m,\Sigma^{\ast})$ with complexity $O(|\Sigma|^3)$ that corrects at most $d^\ast$-errors.

\section{Examples}
Suppose that $p \geq 5$ and let $q=p^e$. Let $E$ be an elliptic curve over $k={\mathbb F}_q$ which is the base extension of a supersingular elliptic curve defined over ${\mathbb F}_p$. The eigenvalues of the $q$-th power Frobenius $F$ are $\{\alpha,\beta\}=\{(\sqrt{-p})^e,(-\sqrt{-p})^e\}$ (\cite{Silverman}) and the characteristic polynomial of $\rho_l(F)$ is $x^2+q$ if $e$ is odd and is $(x+\sqrt{q})^2$ if $e=2f$ where $f$ is odd. 
\begin{fact}(\cite{Zhu}) Let $n$ be a positive odd integer.
\begin{enumerate}
\item Suppose that $p\equiv 1\,({\rm mod}\,4)$ and that $e$ is odd. Then
\[E(k_n)\simeq {\mathbb Z}/(1+q^n){\mathbb Z}.\]
\item Suppose that $e=2f$ where $f$ is odd. Then
\[E(k_n)\simeq {\mathbb Z}/(1+\sqrt{q}^n){\mathbb Z} \oplus {\mathbb Z}/(1+\sqrt{q}^n){\mathbb Z}.\]
\end{enumerate}
\end{fact}

Let $m$ be a positive odd integer. Combine {\bf Lemma 3.1}(2), {\bf Proposition 4.1} and {\bf Fact 6.1} and we obtain the following proposition.
\begin{prop} Let $N$ be a positive odd square free integer. 
\begin{enumerate}
\item Suppose that $p\equiv 1\,({\rm mod}\,4)$ and that $e$ is odd. Then
\[{\rm deg}(D_r)=(-q)_{r},\quad \forall r \in P(N),\]
and
\[|X_N|={\rm deg}(D_N)=1+\sum_{r\in P(N)}\{(-q)_{r}-1\}.\]
\item Suppose that $e=2f$ where $f$ is odd. Then
\[{\rm deg}(D_r)=(\sqrt{q})_{r}^2,\quad \forall r \in P(N), \]
and
\[|X_N|={\rm deg}(D_N)=1+\sum_{r\in P(N)}\{(\sqrt{q})_{r}^2-1\}.\]
\end{enumerate}
\end{prop} 
We fix an integer $m$ greater than $|X_N|$ and let $Z$, $W$ and $\Delta$ be divisors of $E^m$ appeared in the proof of {\bf Proposition 2.2}.
\begin{prop} Let $n$ and $N$ be a positive odd integers satisfying $n>N$ and we assume that $N$ is square free. 
\begin{enumerate}
\item Suppose that $p\equiv 1\,({\rm mod}\,4)$ and that $e$ is odd. Then
\[|E(k_n)^m\setminus (Z\cup W \cup \Delta)| \geq (1+q^n)^{m-1}[1+q^n-m-{m\choose {2}}-\sum_{r\in P(N)} \{m((-q)_r-1)+ {m\choose {(-q)_r}}\}].\]

\item Suppose that $e=2f$ where $f$ is odd. Then
\[|E(k_n)^m\setminus (Z\cup W \cup \Delta)| \geq (1+\sqrt{q}^n)^{2(m-1)}[(1+\sqrt{q}^n)^2-m-{m\choose {2}}-\sum_{r\in P(N)} \{m((\sqrt{q})_r^2-1)+ {m\choose {(\sqrt{q})^2_r}}\}].\]

\end{enumerate}
\end{prop}
{\bf Proof.} (1) We follow the notation of the proof of {\bf Proposition 2.2}. By definition
\[|W\cap E(k_n)^m| \leq \sum_{i=1}^m |\pi_i^{-1}(X_N) \cap E(k_n)^m| \leq m |X_N|\cdot|E(k_n)|^{m-1},\]
and 
\[|W\cap E(k_n)^m| \leq m(1+\sum_{r\in P(N)}\{(-q)_{r}-1\})(1+q^n)^{m-1},\]
by {\bf Fact 6.1} and {\bf Proposition 6.1}. A simple observation shows
\[|\Delta \cap E(k_n)^m| \leq {m\choose{2}}|E(k_n)|^{m-1}={m\choose{2}}(1+q^n)^{m-1}.\]
Let $r\in P(N)$ and $J\in {\mathcal J}_{|X_r|}$. Since $X_r$ is a subset of $E(k_n)$, $\sigma_J$ is an affine surjective map from $E(k_n)^m$ to $E(k_n)$ that is the translation of an surjective group homomorphism by $-\sum_{x\in X_r}x$. Hence
\[|\sigma_J^{-1}(0)|=|E(k_n)|^{m-1}=(1+q^n)^{m-1}.\]
Since $|X_r|=(-q)_r$ by {\bf Proposition 6.1}(1),
\[|Z\cap E(k_n)^m| \leq (1+q^n)^{m-1}\sum_{r\in P(N)} {m\choose{(-q)_r}}.\]
Therefore
\[
|(\Delta\cup Z \cup W)\cap E(k_n)^m|  \leq  (1+q^n)^{m-1}[{m\choose{2}}+m+\sum_{r\in P(N)}\{m((-q)_{r}-1)+{m\choose{(-q)_r}}\}],
\]
and the claim is obtained. One can prove (2) by the similar way.

\begin{flushright} 
$\Box$
\end{flushright}
The proof of {\bf Proposition 2.2} yields the following theorem.
\begin{thm} Let $N$ be a positive odd square free integer and $m$ an integer as below. If an integer $n$ greater than $N$ satisfies one of the following conditions, there is a subset $\Sigma$ of $E(k_n)$ disjoint from ${\rm Spt}(D_N)$ with $|\Sigma|=m$ such that ${\mathcal C}_L^0(D_N,\Sigma^\ast)$ is MDS and that the decomposition ${\mathcal L}_0(D_N)=\oplus_{r\in P(N)}{\mathcal L}_0(D_r)$ is proper.
\begin{enumerate}
\item Suppose that $p\equiv 1\,({\rm mod}\,4)$ and that $e$ is odd. Let $m$ be an integer greater than $1+\sum_{r\in P(N)}\{(-q)_{r}-1\}$. Then
\[q^n>m+{m\choose {2}}+\sum_{r\in P(N)} \{m((-q)_r-1)+ {m\choose {(-q)_r}}\}-1.\]
\item Suppose that $e=2f$ where $f$ is odd. Let $m$ be an integer greater than $1+\sum_{r\in P(N)}\{(\sqrt{q})_{r}^2-1\}$. Then
\[(1+\sqrt{q}^n)^2>m+{m\choose {2}}+\sum_{r\in P(N)} \{m((\sqrt{q})_r^2-1)+ {m\choose {(\sqrt{q})^2_r}}\}.\]
\end{enumerate}
\end{thm}
{\bf Acknowledgement}: This research is partially supported by JSPS Grant in Aid  No. 22540068.

\end{document}